\documentclass[12pt]{article}
\usepackage{amsmath}
\usepackage{amssymb}

\input{tcilatex}
\begin{document}

\begin{center}
\smallskip \ 

\textbf{PHIROTOPES, SUPER P-BRANES AND QUBIT THEORY}

\textbf{\ }

\smallskip \ 

J. A. Nieto \footnote{%
niet@uas.edu.mx,janieto1@asu.edu}

\smallskip

\textit{Facultad de Ciencias F\'{\i}sico-Matem\'{a}ticas, Universidad Aut%
\'{o}noma}

\textit{de Sinaloa, C. P. 80000, Culiac\'{a}n Sinaloa, M\'{e}xico}

\bigskip \ 

\bigskip \ 

\textbf{Abstract}
\end{center}

The phirotope is a complex generalization of the concept of chirotope in
oriented matroid theory. Our main goal in this work is to establish a link
between phirotopes, super $p$-branes and qubit theory. For this purpose we
first discuss maximally supersymmetric solutions of $11$-dimensional
supergravity from the point of view of the oriented matroid theory. We also
clarify a possible connection between oriented matroid theory and
supersymmetry via the Grassmann-Pl\"{u}cker relations. These links are in
turn useful for explaining how our approach can be connected with qubit
theory.

\bigskip \ 

\bigskip \ 

\bigskip \ 

\bigskip \ 

\bigskip \ 

\bigskip \ 

\bigskip \ 


Pacs numbers: 04.60.-m, 04.65.+e, 11.15.-q, 11.30.Ly

March 31, 2014

\newpage \noindent \textbf{1.- Introduction}

\smallskip \ 

Oriented matroid theory [1] is a combinatorial structure that has been
proposed as the underlying mathematical framework for $M$-theory [2]. There
are a number of evidences that suggests that this may be the case, including
the following connections with oriented matroid theory: $p$-branes, qubit
theory, Chern-Simons theory, supergravity and string theory, among others
(see Refs. [3]-[10] and references therein). The key concept to realize
these developments is the so called chirotope notion which provides with one
of the possible axiomatization for oriented matroid theory (see Ref. [1] and
references therein). Since supersymmetry is part of $M$-theory one may
extend such analysis to include complex structure. It turns out that when
the chirotopes are combined with a complex structure one is lead to the
phirotope concept [11]-[13]. Thus, one should expect that when the complex
structure is considered, the link between chirotopes and $p$-branes may be
generalized to a connection between phirotopes and super $p$-branes.

In order to achieve our goal we first explain how phirotopes can be linked
to supersymmetry (see Ref. [9]). In this case, we explain how maximally
supersymmetric solutions of $11$-dimensional supergravities [14]-[15] may be
the key route to construct such a link. This is because the $4$-form $F=dA$
or $F^{\hat{\mu}\hat{\nu}\hat{\alpha}\hat{\beta}}$, with $\hat{\mu},\hat{\nu}%
=0,...,10$, of $11$-dimensional supergravity satisfies the Grassmann-Pl\"{u}%
cker relations (see Ref. [9] and references therein) which in turn are used
to define both chirotopes and phirotopes. In order to clarify this
constructions we briefly review of maximally supersymmetric solution. In
particular, we focus in the algebraic identities of Englert solution [16] of 
$11$-dimensional supergravity. We mention that no only in the case of the
Freund-Rubin solution [17] of $11$-dimensional supergravity the $4$-form
field $F^{\hat{\mu}\hat{\nu}\hat{\alpha}\hat{\beta}}$ admits an
interpretation of a chirotope, but also the Englert solution [16]. In fact,
if one assumes that the only non-vanishing components of $F^{\hat{\mu}\hat{%
\nu}\hat{\alpha}\hat{\beta}}$ are proportional to the completely
antisymmetric symbol $\varepsilon ^{\mu \nu \alpha \beta }$, with $\mu ,\nu
,\alpha ,\beta =0,...,3$, then the Freund-Rubin solution arises from the
bosonic sector of $11$-dimensional supergravity field equations. While, if
in addition, one assumes non-vanishing values for $F^{ijkl}$, with $%
ijkl=4,...,10$, one obtains a the Englert solution. From this perspective it
becomes evident that it is important to study, deeply, the algebraic
properties of $F^{\hat{\mu}\hat{\nu}\hat{\alpha}\hat{\beta}}$. One observes,
for instance, that since in the case of maximally supersymmetric solutions $%
F^{\hat{\mu}\hat{\nu}\hat{\alpha}\hat{\beta}}$ is totally decomposable, it
must be possible to relate $F^{\hat{\mu}\hat{\nu}\hat{\alpha}\hat{\beta}}$
with the chirotope concept via the Grassmann-Pl\"{u}cker relations (see Ref.
[18] for details).

It turns out that a natural generalization of the concept of chirotope is
the so called phirotopes (see Ref. [11]-[13]). The main difference is that
while the chirotope take values in the set $\{-1,0,1\}$ the phirotopes take
values in the set $\{e^{i\theta }\mid 0<\theta <2\pi \}.$ This means that
the phirotopes describe a complex structure. Thus, in principle one can use
phirotopes to introduce grassmann variables and in this way to define the
concept of superphirotope which in turn can be used to establish a link with
super $p$-branes.

The above scenario can be linked with class of $N$-qubits (see Refs. [4]-[5]
and references therein), with the Hilbert space in the form $%
C^{2^{N}}=C^{L}\otimes C^{l}$, with $L=2^{N-n}$ and $l=2^{n}$. In fact, such
a partition allows a geometric interpretation in terms of the complex
Grassmannian variety $Gr(L,l)$ of $l$-planes in $C^{L}$ via the Pl\"{u}cker
embedding [19]. In the case of $N$-rebits one can set a $L\times l$ matrix
variable $b_{a}^{\mu }$, $\mu =1,2,...,L$, $a=1,2...,l$, of $2^{N}=L\times l$
associated with the variable $b_{a_{1}a_{2}...a_{N}}$, with $%
a_{1},a_{2,...}etc$ taking values in the set $\{1,2\}$. Moreover, one can
consider that the first $N-n$ terms in $b_{a_{1}a_{2}...a_{N}}$ are
represented by the index $\mu $ in $b_{a}^{\mu }$, while the remaining $n$
terms are label by the index $a$ in $b_{a}^{\mu }$. One of the advantage of
this construction is that the Pl\"{u}cker coordinates associated with the
real Grassmannians $b_{a}^{\mu }$ are natural invariants of the theory.
Since oriented matroid theory leads to the chirotope concept which is also
defined in terms Pl\"{u}cker coordinates these developments establishes a
possible link between chirotopes and $p$-branes with qubit theory.

This article is organized as follows. In section 2, we present a proof that
a $p$-form is totally decomposable if and only if satisfies the Grassmann-Pl%
\"{u}cker relations. In section 3, Figueroa-O'Farrill-Papadopoulos formalism
of $11$-supergravity is revisited. In section 4, Englert solution of $11$%
-dimensional supergravity is reviewed. In section 5, the chirotope concept
of oriented matroid theory is related to supergravity. In section 6, the
generalization of chirotopes to phirotopes it is discussed. In section 7, we
comment about the relation between maximally supersymmetric solutions of $11$%
-dimensional supergravity and the chirotope concept. In section 8, we
develop the idea of superphirotopes. In section 9, we focus on the a
possible relation between qubit theory and oriented matroid theory. Finally
in section 10, we make some final remarks.

\bigskip \ 

\noindent \textbf{2.- Grassmann-Pl\"{u}cker relations and decomposable }$%
\mathbf{p}$\textbf{-forms}

\smallskip \ 

It is known that the Grassmann-Pl\"{u}cker relation [20] is one of the key
concepts in oriented matroid theory [1]. In order to better understand this
notion it is first convenient to recall the mathematical definition of a
Grassmannian $Gr(p,n)$ (Grassmann variety) over the real $R$ (or any other
field $K$). Let $V$ a vector space of dimensions $n$. The space $Gr(p,n)$
over $R$ is defined as the set of all $p$-dimensional subspaces of $V$.

Here, we are interested in considering the Pl\"{u}cker embedding of $Gr(p,n)$
into the projective space $P(\Lambda ^{p}V)$. Given a subspace $W\in Gr(p,n)$
with basis $\{F^{1},F^{2},...,F^{3}\}$ let a map $f$ be given by%
\begin{equation}
f:W\longrightarrow F^{1}\wedge F^{2}\wedge ...\wedge F^{p},  \tag{1}
\end{equation}%
where the symbol $\wedge $ denotes wedge product. It not difficult to show
that up to scalar multiplication, this map (called Pl\"{u}cker map) is
injective and unique.

It is worth mentioning that, when one is classifying oriented bundles, the
Grassmannian $Gr(p,n)$ can also be denoted by the coset space [20]

\begin{equation}
Gr(p,n)=\frac{SO(n)}{SO(n-p)SO(p)}.  \tag{2}
\end{equation}%
It is interesting to compare (2) with the definition of the $(n-1)$-sphere $%
S^{n-1}$ in terms of the orthogonal group $SO(n)$, namely

\begin{equation}
S^{n-1}=\frac{SO(n)}{SO(n-1)}.  \tag{3}
\end{equation}%
Comparing (2) and (3) one sees that $Gr(p,n)$ is a generalization of $%
S^{n-1} $. Moreover, one may compute the dimension of $Gr(p,n)$ by simply
recalling how is computed the dimension of any coset space $\frac{G}{H}$,
with $H$ a subgroup of $G$. One has $dim\frac{G}{H}=dimG-dimH$. Since $%
dimSO(n)=\frac{n(n-1)}{2}$ one finds the result $dimGr(p,n)=p(n-p)$.

A $p$-form $F_{\mu _{1}...\mu _{p}}\in \Lambda ^{p}V$ is totally
decomposable if there exit a basis $F^{1},...,F^{p}$ such that

\begin{equation}
F_{\mu _{1}...\mu _{p}}\longrightarrow F^{1}\wedge ...\wedge F^{p}.  \tag{4}
\end{equation}%
In order to connect this definition with the Grassmannian $Gr(p,n)$ we first
write%
\begin{equation}
F=\frac{1}{p!}F_{\mu _{1}\mu _{2}...\mu _{p}}e^{\mu _{1}}\wedge e^{\mu
_{2}}\wedge ...\wedge e^{\mu _{p}}.  \tag{5}
\end{equation}%
The expression $e^{\mu _{1}}\wedge e^{\mu _{2}}\wedge ...\wedge e^{\mu _{p}}$
denotes a basis of $\Lambda ^{p}V$. Similarly, one has

\begin{equation}
F^{1}\wedge ...\wedge F^{p}=\frac{1}{p!}\varepsilon
_{a_{1}a_{2}...a_{p}}F_{\mu _{1}}^{a_{1}}F_{\mu _{2}}^{a_{2}}...F_{\mu
_{p}}^{a_{p}}e^{\mu _{1}}\wedge e^{\mu _{2}}\wedge ...\wedge e^{\mu _{p}}. 
\tag{6}
\end{equation}%
The $\varepsilon $-symbol $\varepsilon _{a_{1}a_{2}...a_{p}}$ in (6) is a
completely antisymmetric tensor associated with the $p$-subspace. So, in
this context (4) means that

\begin{equation}
F_{\mu _{1}...\mu _{p}}=\varepsilon _{a_{1}a_{2}...a_{p}}F_{\mu
_{1}}^{a_{1}}F_{\mu _{2}}^{a_{2}}...F_{\mu _{p}}^{a_{p}}.  \tag{7}
\end{equation}%
This result implies that the Pl\"{u}cker map can also be understood by the
transition

\begin{equation}
F_{\mu }^{a}\longrightarrow \varepsilon _{a_{1}a_{2}...a_{p}}F_{\mu
_{1}}^{a_{1}}F_{\mu _{2}}^{a_{2}}...F_{\mu _{p}}^{a_{p}},  \tag{8}
\end{equation}%
where $F_{\mu }^{a}\in Gr(p,n)$.

Now, one may ask: when a $p$-form $F_{\mu _{1}...\mu _{p}}$ is totally
decomposable? There are several ways to approach this question. For
instance, one may prove that $F_{\mu _{1}...\mu _{p}}$ is totally
decomposable if and only if the dimension of all the $v\in V$ dividing $F\in
\Lambda ^{p}(V)$ is $p$ [21]. Here, however, we shall be interested to
consider the Grassmann-Pl\"{u}cker relation

\begin{equation}
F_{\mu _{1}...[\mu _{p}}F_{\nu _{1}...\nu _{p}]}=F_{\mu _{1}...\mu
_{p-1}\alpha _{p+1}}F_{\alpha _{1}...\alpha _{p}}\delta _{\nu _{1}...\nu
_{p}\mu _{p}}^{\alpha _{1}...\alpha _{p}\alpha _{p+1}}=0.  \tag{9}
\end{equation}%
Here, the symbol $\delta _{\nu _{1}...\nu _{p}\mu _{p}}^{\alpha
_{1}...\alpha _{p}\alpha _{p+1}}$ denotes a generalized delta. The idea is
now to prove that a $p$-form $F_{\mu _{1}...\mu _{p}}$ is totally
decomposable if and only if the Grassmann-Pl\"{u}cker relation (9) holds.

If $F_{\mu _{1}...\mu _{p}}$ is totally decomposable then one sees that
using (7) the combination

\begin{equation}
F_{\mu _{1}...[\mu _{p}}F_{\nu _{1}...\nu _{p}]}  \tag{10}
\end{equation}%
leads to

\begin{equation}
\varepsilon _{a_{1}a_{2}...[a_{p}}\varepsilon _{b_{1}b_{2}...b_{p}]}F_{\mu
_{1}}^{a_{1}}F_{\mu _{2}}^{a_{2}}...F_{\mu _{p}}^{a_{p}}F_{\nu
_{1}}^{b_{1}}F_{\nu _{2}}^{b_{2}}...F_{\nu _{p}}^{b_{p}}.  \tag{11}
\end{equation}%
But one has%
\begin{equation}
\varepsilon _{a_{1}a_{2}...[a_{p}}\varepsilon _{b_{1}b_{2}...b_{p}]}\equiv 0.
\tag{12}
\end{equation}%
So, if the Grassmann-Pl\"{u}cker relation (9) holds then $F_{\mu _{1}...\mu
_{p}}$ is totally decomposable. Perhaps, it is more difficult to prove that
(9) implies (7). This can be shown using an induction method (see [21] and
references therein), but here we present an alternative prove that we are
not aware of its existence in the literature.

Let $F_{\mu }^{A}$ be an extended basis of $V$. We can define

\begin{equation}
F^{A_{1}...A_{p}}\equiv F^{\mu _{1}...\mu _{p}}F_{\mu _{1}}^{A_{1}}...F_{\mu
_{p}}^{A_{p}}.  \tag{13}
\end{equation}%
Considering the inverse $F_{A}^{\mu }$ of $F_{\mu }^{A}$ this expression
leads to

\begin{equation}
F_{\mu _{1}...\mu _{p}}=F_{A_{1}...A_{p}}F_{\mu _{1}}^{A_{1}}...F_{\mu
_{p}}^{A_{p}}.  \tag{14}
\end{equation}%
Using (13), it is not difficult to see that (9) implies

\begin{equation}
F_{A_{1}...[A_{p}}F_{B_{1}...B_{p}]}=F_{A_{1}...A_{p-1}C_{p+1}}F_{C_{1}...C_{p}}\delta _{B_{1}...B_{p}A_{p}}^{C_{1}...C_{p}C_{p+1}}=0.
\tag{15}
\end{equation}%
Assume that (15) holds. Let us apply (15) to the particular case%
\begin{equation}
F_{a_{1}...a_{p-1}C_{p+1}}F_{C_{1}...C_{p}}\delta
_{b_{1}...b_{p}A_{p}}^{C_{1}...C_{p}C_{p+1}}=0,  \tag{16}
\end{equation}%
with $C_{p}\neq a$ and $a$ and $b$ running in the dimension of the $p$%
-subspace. One can show that (16) leads to%
\begin{equation}
F_{a_{1}...a_{p-1}A_{p}}F_{b_{1}...b_{p}}=0.  \tag{17}
\end{equation}%
Since in general%
\begin{equation}
F_{b_{1}...b_{p}}=\Lambda \varepsilon _{b_{1}b_{2}...b_{p}}\neq 0,  \tag{18}
\end{equation}%
with $\Lambda $ an arbitrary constant, one finds that

\begin{equation}
F_{a_{1}...a_{p-1}A_{p}}=0.  \tag{19}
\end{equation}%
Now, considering the next particular case%
\begin{equation}
F_{a_{1}...a_{p-2}C_{p+2}C_{p+1}}F_{C_{1}...C_{p}}\delta
_{b_{1}...b_{p}A_{p-1}A_{p}}^{C_{1}...C_{p}C_{p+1}C_{p+2}}=0  \tag{20}
\end{equation}%
and using (19) one obtains

\begin{equation}
F_{a_{1}...a_{p-2}A_{p-1}A_{p}}=0.  \tag{21}
\end{equation}%
Following similar procedure one ends up with that the result that the only
non vanishing components of $F_{A_{1}...A_{p}}$ are given by

\begin{equation}
F_{a_{1}a_{2}...a_{p}}\neq 0.  \tag{22}
\end{equation}%
But, one knows that $F_{a_{1}a_{2}...a_{p}}=\Lambda \varepsilon
_{a_{1}a_{2}...a_{p}}$. Therefore, using (14) we obtain

\begin{equation}
F_{\mu _{1}\mu _{2}...\mu _{p}}=\Lambda \varepsilon
_{a_{1}a_{2}...a_{p}}F_{\mu _{1}}^{a_{1}}F_{\mu _{2}}^{a_{2}}...F_{\mu
_{p}}^{a_{p}}.  \tag{23}
\end{equation}%
Up to constant, this expression corresponds to (7) meaning that $F_{\mu
_{1}\mu _{2}...\mu _{p}}$ is decomposable. The expression (23) will be very
useful in the next sections.

\bigskip \ 

\bigskip \ 

\bigskip \ 

\bigskip \ 

\noindent \textbf{3.- Figueroa-O'Farrill-Papadopoulos formalism revisited}

\smallskip \ 

Consider the $4$-form $F_{\mu _{1}\mu _{2}\mu _{3}\mu _{4}}$. We shall
assume that this form satisfies the Grassmann Pl\"{u}cker relation%
\begin{equation}
F_{\mu _{1}\mu _{2}\mu _{3}[\mu _{4}}F_{\nu _{1}\nu _{2}\nu _{3}\nu _{4}]}=0.
\tag{24}
\end{equation}%
It turns out that (24) holds if any only if the following two the relations
are satisfied

\begin{equation}
F_{[\mu _{1}\mu _{2}\mu _{3}\mu _{4}}F_{\nu _{1}\nu _{2}\nu _{3}\nu _{4}]}=0,
\tag{25}
\end{equation}%
and

\begin{equation}
F_{\mu _{1}[\mu _{2}\mu _{3}\mu _{4}}F_{\nu _{1}\nu _{2}\nu _{3}]\nu _{4}}=0.
\tag{26}
\end{equation}%
It is worth mentioning that (25) and (26) play a crucial role in maximally
supersymmetric $11$-dimensional supergravity [14]-[15]. Let us prove that in
fact this result holds. First, one observes that in general the bracket $[,]$
in (24)-(26) can be written as

\begin{equation}
G_{[\mu _{1}...\mu _{d+1}]}\equiv G_{\alpha _{1}...\alpha _{d+1}}\delta
_{\mu _{1}...\mu _{d+1}}^{\alpha _{1}...\alpha _{d+1}}.  \tag{27}
\end{equation}%
The quantity $G_{\alpha _{1}...\alpha _{d+1}}$ is any $d+1$-rank tensor.
Considering the fact that%
\begin{equation}
\delta _{\mu _{1}...\mu _{d+1}}^{\alpha _{1}...\alpha _{d+1}}=\delta _{\mu
_{1}}^{\alpha _{1}}\delta _{\mu _{2}...\mu _{d+1}}^{\alpha _{2}...\alpha
_{d+1}}+\tsum \limits_{k=2}^{d+1}(-1)^{k}\delta _{\mu _{k}}^{\alpha
_{1}}\delta _{\mu _{2}...\hat{\mu}_{k}...\mu _{d+1}}^{\alpha _{2}...\alpha
_{d+1}},  \tag{28}
\end{equation}%
where $\hat{\mu}_{k}$ means omitting this index, one finds that (25) follows
if and only if one has

\begin{equation}
F_{\mu _{1}\alpha _{2}\alpha _{3}\alpha _{4}}F_{\beta _{1}\beta _{2}\beta
_{3}\beta _{4}}\delta _{\mu _{2}\mu _{3}\mu _{4}\nu _{1}\nu _{2}\nu _{3}\nu
_{4}}^{\alpha _{2}\alpha _{3}\alpha _{4}\beta _{1}\beta _{2}\beta _{3}\beta
_{4}}=0,  \tag{29}
\end{equation}%
which means%
\begin{equation}
F_{\mu _{1}[\mu _{2}\mu _{3}\mu _{4}}F_{\nu _{1}\nu _{2}\nu _{3}\nu _{4}]}=0.
\tag{30}
\end{equation}

Properly applying again (28) one gets

\begin{equation}
F_{\mu _{1}[\mu _{2}\mu _{3}\mu _{4}}F_{\nu _{1}\nu _{2}\nu _{3}\nu
_{4}]}=3F_{\mu _{1}\mu _{2}[\mu _{3}\mu _{4}}F_{\nu _{1}\nu _{2}\nu _{3}\nu
_{4}]}+4F_{\mu _{1}[\mu _{3}\mu _{4}\nu _{1}}F_{\nu _{2}\nu _{3}\nu _{4}]\mu
_{2}}.  \tag{31}
\end{equation}%
Thus, considering the fact that the (26) holds the first term in (31)
vanishes, that is%
\begin{equation}
F_{\mu _{1}\mu _{2}[\mu _{3}\mu _{4}}F_{\nu _{1}\nu _{2}\nu _{3}\nu _{4}]}=0.
\tag{32}
\end{equation}

Similar technique it leads us to the identity

\begin{equation}
F_{\mu _{1}\mu _{2}[\mu _{3}\mu _{4}}F_{\nu _{1}\nu _{2}\nu _{3}\nu
_{4}]}=2F_{\mu _{1}\mu _{2}\mu _{3}[\mu _{4}}F_{\nu _{1}\nu _{2}\nu _{3}\nu
_{4}]}+4F_{\mu _{1}\mu _{2}[\mu _{4}\nu _{1}}F_{\nu _{2}\nu _{3}\nu _{4}]\mu
_{3}}.  \tag{33}
\end{equation}%
which in turn gives,

\begin{equation}
F_{\mu _{1}\mu _{2}\mu _{3}[\mu _{4}}F_{\nu _{1}\nu _{2}\nu _{3}\nu
_{4}]}=-2F_{\mu _{1}\mu _{2}[\mu _{4}\nu _{1}}F_{\nu _{2}\nu _{3}\nu
_{4}]\mu _{3}}.  \tag{34}
\end{equation}%
This expression implies that the right hand side of (34) is antisymmetric in
the indices $\mu _{1}$ and $\mu _{3}$.

On the other hand one obtains

\begin{equation}
\begin{array}{c}
F_{\mu _{1}[\mu _{2}\mu _{4}\nu _{1}}F_{\nu _{2}\nu _{3}\nu _{4}]\mu
_{3}}=3F_{\mu _{1}\mu _{2}[\mu _{4}\nu _{1}}F_{\nu _{2}\nu _{3}\nu _{4}]\mu
_{3}}-3F_{\mu _{1}[\mu _{4}\nu _{1}\nu _{2}}F_{\nu _{3}\nu _{4}]\mu _{2}\mu
_{3}} \\ 
\\ 
=3F_{\mu _{1}\mu _{2}[\mu _{4}\nu _{1}}F_{\nu _{2}\nu _{3}\nu _{4}]\mu
_{3}}-3F_{\mu _{3}\mu _{2}[\mu _{4}\nu _{1}}F_{\nu _{2}\nu _{3}\nu _{4}]\mu
_{1}}.%
\end{array}
\tag{35}
\end{equation}%
From (26) one sees that the left hand side of (35) vanishes and therefore we
obtain

\begin{equation}
F_{\mu _{1}\mu _{2}[\mu _{4}\nu _{1}}F_{\nu _{2}\nu _{3}\nu _{4}]\mu
_{3}}=F_{\mu _{3}\mu _{2}[\mu _{4}\nu _{1}}F_{\nu _{2}\nu _{3}\nu _{4}]\mu
_{1}}.  \tag{36}
\end{equation}%
This means that $F_{\mu _{1}\mu _{2}[\mu _{4}\nu _{1}}F_{\nu _{2}\nu _{3}\nu
_{4}]\mu _{3}}$ is symmetric in the indices $\mu _{1}$ and $\mu _{3}$ which
contradicts the conclusion below (34). Thus, we have found that the only
consistent possibility is to set

\begin{equation}
F_{\mu _{1}\mu _{2}[\mu _{4}\nu _{1}}F_{\nu _{2}\nu _{3}\nu _{4}]\mu _{3}}=0,
\tag{37}
\end{equation}%
which implies (24) via (34). Summarizing, we have shown that (25) and (26)
imply (24) which is the Grassmann-Pl\"{u}cker relation. Conversely, using
once again the properties of the generalized delta $\delta _{\mu _{1}...\mu
_{d+1}}^{\alpha _{1}...\alpha _{d+1}}$ one can show that both $F_{\mu
_{1}[\mu _{2}\mu _{3}\mu _{4}}F_{\nu _{1}\nu _{2}\nu _{3}\nu _{4}]}$ and $%
F_{\mu _{1}[\mu _{3}\mu _{4}\nu _{1}}F_{\nu _{2}\nu _{3}\nu _{4}]\mu _{2}}$
can be written in terms of $F_{\mu _{1}\mu _{2}\mu _{3}[\mu _{4}}F_{\nu
_{1}\nu _{2}\nu _{3}\nu _{4}]}$ and therefore (24) implies (25) and (26).
This means that the expression (24) is equivalent to the two formulae (25)
and (26). Thus, we have complete an alternative proof of such a equivalence.

The formula (24) implies that $F_{\mu _{1}\mu _{2}\mu _{3}\mu _{4}}$ is
totally decomposable. This means that there exist $(4\times d+1)$-matrices $%
F_{a}^{\mu }$ in $Gr(p,n)$ such that $F_{\mu _{1}\mu _{2}\mu _{3}\mu _{4}}$
can be written in the form

\begin{equation}
F^{\mu _{1}\mu _{2}\mu _{3}\mu _{4}}=\varepsilon
^{a_{1}a_{2}a_{3}a_{4}}F_{a_{1}}^{\mu _{1}}F_{a_{2}}^{\mu
_{2}}F_{a_{3}}^{\mu _{3}}F_{a_{4}}^{\mu _{4}}.  \tag{38}
\end{equation}%
Thus, one may conclude that maximally supersymmetric solutions of $11$%
-dimensional supergravity implies that $F^{\mu _{1}\mu _{2}\mu _{3}\mu _{4}}$
can be written as (38).

It turns out convenient to briefly mention how the above result is linked to
maximally supersymmetric solution of $11$-dimensional supergravity. In fact,
Figueroa-O'Farrill and Papadopoulos proved that such a solution must be
isometric to either $AdS_{4}\times S^{7}$ or $AdS_{7}\times S^{4}$. Their
starting point in this result is the vanishing of the curvature $\mathcal{R}$
of the supercovariant connection $\mathcal{D}$ living in $(M^{11},g,F)$. In
fact, demanding the vanishing of the curvature $\mathcal{R}$ they found that 
$(M^{11},g,F)$ is maximally supersymmetric solution if and only if $%
(M^{11},g)$ is locally symmetric space and $F$ is parallel and decomposable.

Let us clarify further this theorem. In the non-degenerate case, spontaneous
compactification allows to assume that the only nonvanishing components of $%
F_{a}^{\mu }$ are $F_{a}^{\mu }\sim \delta _{a}^{\mu }$, with $\mu =0,1,2,3$
or $F_{a}^{\hat{\mu}}\sim F_{a}^{\hat{\mu}}$, with $\hat{\mu}=8,9,10,11$
leading to the two possible solutions $AdS_{4}\times S^{7}$ or $%
AdS_{7}\times S^{4}$, respectively. In fact, in the first case one gets that
the only nonvanishing components of $F^{\mu _{1}\mu _{2}\mu _{3}\mu _{4}}$
are $F^{\mu \nu \alpha \beta }\sim \varepsilon ^{\mu \nu \alpha \beta }$.
Thus, as seen from the $11$-dimensional field equations

\begin{equation}
\begin{array}{c}
\frac{1}{3!}\varepsilon _{\mu _{1}\mu _{2}\mu _{3}\mu _{4}\nu _{1}\nu
_{2}\nu _{3}\nu _{4}NPQ}F^{NPQM};_{M}=\frac{1}{2(4!)^{2}}F_{[\mu _{1}\mu
_{2}\mu _{3}\mu _{4}}F_{\nu _{1}\nu _{2}\nu _{3}\nu _{4}]}, \\ 
\\ 
R_{MN}-\frac{1}{2}g_{MN}R=\frac{1}{6}F_{MPQR}F_{N}^{PQR}-\frac{1}{48}%
g_{MN}F_{SPQR}F^{SPQR},%
\end{array}
\tag{39}
\end{equation}%
one obtains the Freund-Rubin solution $AdS_{4}\times S^{7}$. While in the
second case one assumes the solution $F^{\hat{\mu}\hat{\nu}\hat{\alpha}\hat{%
\beta}}\sim \varepsilon ^{\hat{\mu}\hat{\nu}\hat{\alpha}\hat{\beta}}$ and
the field equations (39) lead to the solution $AdS_{7}\times S^{4}.$

Perhaps, it is also convenient to write the three equations (24)-(26) in
abstract notation. From the formula $\mathcal{R}=0$ one can essentially
derive two algebraic formulae for $F$ (25) and (26) which in abstract
notation become

\begin{equation}
F\wedge F=0  \tag{40}
\end{equation}%
and

\begin{equation}
_{\iota _{X}}F\wedge _{\iota _{Y}}F=0,  \tag{41}
\end{equation}%
respectively. Here $\iota _{X}$ and $\iota _{Y}$ denote an inner product for
the two arbitrary vectors $X$ and $Y$, respectively. From (25) and (26) we
proved that $F$ satisfies (24) which in abstract notation is written as

\begin{equation}
_{\iota _{Z}\iota _{Y}\iota _{X}}F\wedge F=0.  \tag{42}
\end{equation}

It is interesting to mention the way that Figueroa-O'Farrill and
Papadopoulos prove that (25) and (26) imply (24). They first observe that
contracting (25) with respect to the three vectors $X,Y$ and $Z$ one obtains

\begin{equation}
_{\iota _{Z}\iota _{Y}\iota _{X}}F\wedge F=-_{\iota _{Y}\iota _{X}}F\wedge
_{\iota _{Z}}F.  \tag{43}
\end{equation}%
While, contracting equation (26) with a third vector field one gets

\begin{equation}
_{\iota _{Y}\iota _{X}}F\wedge _{\iota _{Z}}F=_{\iota _{Y}\iota _{Z}}F\wedge
_{\iota _{X}}F.  \tag{44}
\end{equation}%
Thus, comparing (43) and (44) one sees that whereas (42) implies that the
expression $_{\iota _{Y}\iota _{X}}F\wedge _{\iota _{Z}}F$ is symmetric in $%
X $ and $Z$, (44)\ means that it is skew-symmetric. This means that the term 
$_{\iota _{Y}\iota _{X}}F\wedge _{\iota _{Z}}F$ must vanish and therefore
(42) follows (see Refs. [14] and [15] for details).

\bigskip \ 

\noindent \textbf{4.- Englert solution revisited}

\smallskip \ 

Consider the octonionic identity [22],

\begin{equation}
f^{ijkl}f_{mnrl}=\delta _{m}^{[i}\delta _{n}^{j}\delta _{r}^{k]}+\frac{1}{4}%
f_{[mn}^{[ij}\delta _{r]}^{k]},  \tag{45}
\end{equation}%
with the indices $i,j,...etc$ running from $4$ to $11$. Here, $f_{ijkl}$ is
a self dual object. Furthermore, $f_{ijkl}$ is defined in terms of the
octonionic structure constants $\psi _{ijk}$ and its dual $\varphi _{ijkl}$
through the relations

\begin{equation}
f_{ijk11}=\psi _{ijk}  \tag{46}
\end{equation}%
and

\begin{equation}
f_{ijkl}=\varphi _{ijkl}.  \tag{47}
\end{equation}%
From (45) it is not difficult to see that

\begin{equation}
f_{[ijk}^{r}f_{lmn]r}=0.  \tag{48}
\end{equation}%
This expression can be understood as a solution for

\begin{equation}
f_{s[ijk}f_{lmn]r}=0,  \tag{49}
\end{equation}%
which remains us the formula (26) reduced to seven dimensions. In fact,
introducing a sieben-bein $h_{k}^{i}$ one can make this identification more
transparent [22]. In fact, one has

\begin{equation}
F_{ijkl}=h_{i}^{r}h_{j}^{s}h_{k}^{t}h_{l}^{m}f_{rstm}  \tag{50}
\end{equation}%
and therefore (49) leads to

\begin{equation}
F_{s[ijk}F_{lmn]r}=0.  \tag{51}
\end{equation}%
Starting from (45) and following similar arguments we may establish that

\begin{equation}
F_{s[ijk}F_{lmnr]}=0  \tag{52}
\end{equation}%
and

\begin{equation}
F_{[sijk}F_{lmnr]}=0.  \tag{53}
\end{equation}%
Thus, according to the discussion of previous sections (52)\ and (53) imply
that $F_{ijkl}$ satisfies the relation

\begin{equation}
F_{sij[k}F_{lmnr]}=0,  \tag{54}
\end{equation}%
which means that $F_{ijkl}$ is decomposable.

On the other hand, in four dimensions as we already mentioned, we can take

\begin{equation}
F^{\mu \nu \alpha \beta }=\Lambda \varepsilon ^{\mu \nu \alpha \beta }, 
\tag{55}
\end{equation}%
where $\Lambda $ is an arbitrary function. Since $\varepsilon ^{\mu \nu
\alpha \beta }$ is a maximally completely antisymmetric object in four
dimensions we get the formula

\begin{equation}
F_{\mu \nu \alpha \lbrack \beta }F_{\sigma \rho \tau \gamma ]}=0,  \tag{56}
\end{equation}%
which implies

\begin{equation}
F_{[\mu \nu \alpha \beta }F_{\sigma \rho \tau \gamma ]}=0.  \tag{57}
\end{equation}%
Thus, $F^{\mu \nu \alpha \beta }$ is also decomposable.

Our main observation is that despite both $F_{ijkl}$ and $F_{\mu \nu \alpha
\beta }$ are both decomposable, the $11$-dimensional components $F_{A\nu
\alpha D}$ are not. The reason comes from the fact that in spite that $%
F_{ijkl}$ and $F_{\mu \nu \alpha \beta }$ are decomposable the components of 
$F_{A\nu \alpha D}$ not necessarily satisfies the relation $%
F_{A_{1}A_{2}A_{3}[A_{4}}F_{\nu _{1}\nu _{2}\nu _{3}\nu _{4}]}=0$. The
result follows from the expression

\begin{equation}
F_{\mu \nu \alpha \lbrack \beta }F_{ijkm]}\neq 0,  \tag{58}
\end{equation}%
or

\begin{equation}
F_{[\mu \nu \alpha \beta }F_{ijkm]}\neq 0.  \tag{59}
\end{equation}%
So, it turns out that full $F_{ABCD}$ is not decomposable. In fact, since $%
\varepsilon ^{\mu \nu \alpha \beta }$ and $f^{ijkm}$ take values in the set $%
\{-1,0,1\}$ in general we have that

\begin{equation}
\varepsilon _{\mu \nu \alpha \lbrack \beta }f_{ijkm]}\neq 0,  \tag{60}
\end{equation}%
or

\begin{equation}
\varepsilon _{\lbrack \mu \nu \alpha \beta }f_{ijkm]}\neq 0.  \tag{61}
\end{equation}%
In turn this means that $F_{[A_{1}A_{2}A_{3}A_{4}}F_{\nu _{1}\nu _{2}\nu
_{3}\nu _{4}]}\neq 0$ or $F\wedge F\neq 0$. Consequently we no longer have
maximally supersymmetric solution. Nevertheless, as Englert showed, although
the right hand side of the first field equation in (39) is not vanishing the
field equations still admit the solution $AdS_{4}\times S^{7}$. This means
that maximally supersymmetric solutions can be considered as a broken
symmetry (see Ref. [18] and references therein).

\bigskip \ 

\noindent \textbf{5.- Connection with chirotopes}

\smallskip \ 

The aim of this section is to discuss part of the formalism described in
section 2, 3 and 4 from the point of view of the oriented matroid theory.
Indeed, our discussion will focus on the chirotope concept which provides
one possible definition of an oriented matroid [1]. In fact, chirotopes has
been a major subject of investigation in mathematics during the last 25
years [1]. Roughly speaking a chirotope is a combinatorial abstraction of
subdeterminants of a given matrix. More formally, a realizable $p$-rank
chirotope is an alternating function $\chi :\{1,...,n\}^{p}\rightarrow
\{-1,0,1\}$ satisfying the Grassmann-Pl\"{u}cker relation

\begin{equation}
\chi _{\hat{A}_{1}...\hat{A}_{n-1}[\hat{A}_{p}}\chi _{\hat{B}_{1}...\hat{B}%
_{p}]}=0,  \tag{62}
\end{equation}%
while nonrealizable $p$-rank chirotope corresponds to the case

\begin{equation}
\chi _{\hat{A}_{1}...\hat{A}_{n-1}[\hat{A}_{p}}\chi _{\hat{B}_{1}...\hat{B}%
_{p}]}\neq 0.  \tag{63}
\end{equation}%
It is worth mentioning that there is a close connection between chirotopes
and Grassmann variety. In fact, the Grassmann-Pl\"{u}cker relations describe
a projective embedding of the Grassmannian of planes via decomposable $p$%
-forms (see Ref. [1] for details).

Thanks to our revisited review of Freund-Rubin and Englert solutions given
in the previous sections we find that the link between these solutions and
the chirotope is straightforward. In fact, our first observation is that any 
$\varepsilon $-symbol is in fact a realizable chirotope (see Ref. [23]),
since it is always true that

\begin{equation}
\varepsilon _{\hat{A}_{1}...\hat{A}_{p-1}[\hat{A}_{p}}\varepsilon _{\hat{B}%
_{1}...\hat{B}_{p}]}=0.  \tag{64}
\end{equation}%
From this perspective we recognize that the formula (24) indicates that in
the case of maximally supersymmetric solutions, in $11$-dimensional
supergravity, the $4$-form $F_{ABCD}$ is a realizable $4$-rank chirotope.
While in the case of Freund-Rubin-Englert solution, from (39) and (42) one
discovers that according to our discussion of section $4$ one may identify $%
F_{ABCD}$ with a nonrealizable $4$-rank chirotope. From this connections one
may expect that there may be many possible $4$-rank chirotopes in $11$%
-dimensions and therefore there must be many new and unexpected solutions
for $11$-dimensional supergravity.

One of our key tools in our formalism is the octonionic structure. This
division algebra was already related to the Fano matroid and therefore, a
possible connection with supergravity was established (see Ref. [2] and
references therein). Here, we have been more specific and through the
chirotope concept we established the relation between the
Freund-Rubin-Englert solution and oriented matroid theory. However, it may
be interesting to understand the possible role of the Fano matroid in this
scenario.

Moreover, here we focused on $11$-dimensional supergravity but, in
principle, one may expect to apply similar procedure in the case of $10$%
-dimensional supergravity and other higher dimensional supergravities such
as Type I supergravity and massive IIA supergravity.

An important property in the oriented matroid theory is that one can
associate any chirotopes with its dual. Thus, working on the framework of
oriented matroids we can assure that any possible solution for $11$%
-dimensional supergravity in terms of chirotopes will have a dual solution.
This means that this kind of solution contains automatically a dual symmetry.

It is worth mentioning that using the idea of matroid bundle [24]-[28], Guha
[29] has observed that chirotopes can be related to Nambu-Poisson structure.
It may be interesting to see whether this Nambu-Poisson structure is related
to $11$-dimensional supergravity.

\bigskip \ 

\bigskip \ 

\noindent \textbf{6.- Chirotope and Phirotope concepts}

\smallskip \ 

Let us start considering again the completely antisymmetric symbol

\begin{equation}
\varepsilon ^{a_{1}...a_{d}}\in \{-1,0,1\}.  \tag{65}
\end{equation}%
In this section, the indices $a_{1},...,a_{d}$ run from $1$ to $d$. This is
a $d$-rank tensor which values are $+1$ or $-1$ depending on even or odd
permutations of $\varepsilon ^{12...d}$, respectively. Moreover, $%
\varepsilon ^{a_{1}...d_{d}}$ takes the value $0$ unless $a_{1}...a_{d}$ are
all different. Let $v_{a}^{i}$ be any $d\times n$ matrix over some field $F$%
, where the index $i$ takes values in the set $E=\{1,...,n\}$. Consider the
object%
\begin{equation}
\Sigma ^{i_{1}...i_{d}}=\varepsilon
^{a_{1}...a_{d}}v_{a_{1}}^{i_{1}}...v_{a_{d}}^{i_{d}},  \tag{66}
\end{equation}%
which can also be written as%
\begin{equation}
\Sigma ^{i_{1}...i_{d}}=\det (\mathbf{v}^{i_{1}},...,\mathbf{v}^{i_{d}}). 
\tag{67}
\end{equation}%
Using the $\varepsilon $-symbol property

\begin{equation}
\varepsilon ^{a_{1}...[a_{d}}\varepsilon ^{b_{1}...b_{d}]}=0.  \tag{68}
\end{equation}%
It is not difficult to prove that $\Sigma ^{i_{1}...i_{d}}$ satisfies the
Grassmann-Pl\"{u}cker relations, namely

\begin{equation}
\Sigma ^{i_{1}...[i_{d}}\Sigma ^{j_{1}...j_{d}]}=0.  \tag{69}
\end{equation}%
We recall that the brackets in the indices of (68) and (69) mean completely
antisymmetrized.

A realizable chirotope $\chi $ is defined as

\begin{equation}
\chi ^{i_{1}...i_{d}}=sign\Sigma ^{i_{1}...i_{d}}.  \tag{70}
\end{equation}%
From the point of view of exterior algebra one finds that there is a close
connection between Grassmann algebra and a chirotope. Let us denote by $%
\wedge _{d}R^{n}$ the $(_{d}^{n})$-dimensional real vector space of
alternating $d$-forms on $R^{n}$. We recall that an element $\mathbf{\Sigma }
$ in $\wedge _{d}R^{n}$ is said to be decomposable if

\begin{equation}
\mathbf{\Sigma }=\mathbf{v}_{1}\wedge \mathbf{v}_{2}\wedge ...\wedge \mathbf{%
v}_{d},  \tag{71}
\end{equation}%
for some $\mathbf{v}_{1},\mathbf{v}_{2},...,.\mathbf{v}_{d}\in R^{n}$. It is
not difficult to see that (71) can also be written as

\begin{equation}
\mathbf{\Sigma }=\frac{1}{r!}\Sigma ^{i_{1}...i_{d}}e_{i_{1}}\wedge
e_{i_{2}}\wedge ...\wedge e_{i_{d}},  \tag{72}
\end{equation}%
where $e_{i_{1}},e_{i_{2}},...,e_{i_{d}}$ are $1$-form bases in $R^{n}$ and $%
\Sigma ^{i_{1}...i_{d}}$ is given in (66). This shows that $\Sigma
^{i_{1}...i_{d}}$ can be identified with an alternating decomposable $d$%
-form.

In order to define non-realizable chirotopes it is convenient to write the
expression (69) in the alternative form

\begin{equation}
\tsum \limits_{k=1}^{d+1}s_{k}=0,  \tag{73}
\end{equation}%
where

\begin{equation}
s_{k}=(-1)^{k}\Sigma ^{i_{1}...i_{d-1}j_{k}}\Sigma ^{j_{1}...\hat{\jmath}%
_{k}...j_{d+1}}.  \tag{74}
\end{equation}%
Here, $j_{d+1}=i_{d}$ and $\hat{\jmath}_{k}$ establish the notation for
omitting this index. Thus, for a general definition one defines a $d$-rank
chirotope $\chi :E^{d}\rightarrow \{-1,0,1\}$ if there exist $%
r_{1},...,r_{d+1}\in R^{+}$ such that

\begin{equation}
\tsum \limits_{k=1}^{d+1}r_{k}s_{k}=0,  \tag{75}
\end{equation}%
with

\begin{equation}
s_{k}=(-1)^{k}\chi ^{i_{1}...i_{d-1}j_{k}}\chi ^{j_{1}...\hat{\jmath}%
_{k}...j_{d+1}},  \tag{76}
\end{equation}%
and $k=1,...,d+1$. It is evident that (73) is a particular case of (76).
Therefore, there are chirotopes that may be non-realizable. Moreover, this
definition of a chirotope is equivalent to various others (see Refs.
[11]-[13] for details), but it seems that the present one is more convenient
for a generalization to the complex structure setting.

The generalization of a chirotope to a phirotope is straightforward. A
function $\varphi :E^{d}\rightarrow S^{1}\cup \{0\}$ on all $d$-tuples of $%
E=\{1,...,n\}$ is called a $d$-rank phirotope if (a) $\varphi $ is
alternating and (b) for

\begin{equation}
\omega _{k}=(-1)^{k}\varphi ^{i_{1}...i_{d-1}j_{k}}\varphi ^{j_{1}...\hat{%
\jmath}_{k}...j_{d+1}}=0,  \tag{77}
\end{equation}%
for $k=1,...,d+1$ there exist $r_{1},...,r_{d+1}\in R^{+}$ such that

\begin{equation}
\tsum \limits_{k=1}^{d+1}r_{k}\omega _{k}=0.  \tag{78}
\end{equation}

In the case of a realizable phirotope we have

\begin{equation}
\Omega ^{i_{1}...i_{d}}=\omega (\det (\mathbf{u}^{i_{1}},...,\mathbf{u}%
^{i_{d}})),  \tag{79}
\end{equation}%
where $\omega (z)\in S^{1}\cup \{0\}$ and $(\mathbf{u}^{i_{1}}...\mathbf{u}%
^{i_{d}})$ are a set of complex vectors in $C^{d}$. We observe that one of
the main differences between a chirotope and a phirotope is that the image
of a phirotope is no longer a discrete set (see Refs. [11]-[13] for details).

\bigskip \ 

\bigskip \ 

\bigskip \ 

\noindent \textbf{7- Supergravity and phirotopes}

\smallskip \ 

As we mentioned in section 3, maximally supersymmetric solution of $11$%
-dimensional supergravity leads to the two conditions

\begin{equation}
F_{M[L_{1}L_{2}L_{3}}F_{L_{4}L_{5}L_{6}L_{7}]}=0  \tag{80}
\end{equation}%
and

\begin{equation}
F_{M[P_{1}P_{2}P_{3}}F_{Q_{1}Q_{2}Q_{3}]N}=0,  \tag{81}
\end{equation}%
for the $4$-form field strength $F=dA$ which are equivalent to the
Grassmann-Pl\"{u}cker relations

\begin{equation}
F_{MP_{1}P_{2}[P_{3}}F_{Q_{1}Q_{2}Q_{3}Q_{4}]}=0,  \tag{82}
\end{equation}%
meaning that $F$ is decomposable. Thus, according to the discussion of the
previous sections one discovers that (82) establishes that $F$ is a
realizable $4$-rank chirotope with a ground set $E=\{1,...,11\}$. This in
turn means that maximal supersymmetry in $11$-dimensional supergravity is
related to oriented matroid theory. Similar conclusion can be obtained for
the case of $10$-dimensional supergravity. Hence, one may understand the
chirotope concept as the bridge between supersymmetry and the oriented
matroid theory. Thus, one should expect a generalization of oriented matroid
theory which would include supersymmetry. But in order to develop this idea
it turns out more convenient to consider a complex structure, and this means
that we need to focus on the superphirotope notion rather than on the
superchirotope concept which must arise as a particular case of the former.

\bigskip \ 

\noindent \textbf{8. Superphirotope}

\smallskip \ 

The main goal of this section is to outline a possible supersymmetrization
of a phirotope. By convenience we shall call superphirotope such a
supersymmetric phirotope. Inspired in super $p$-brane theory one finds that
one way to define a superphirotope, which assures supersymmetry, is as
follows. First, we need to locally consider the expressions (77)-(79) in the
sense that $\varphi ^{i_{1}....j_{d}}(\xi )$ is a local phirotope if

\begin{equation}
\omega _{k}=(-1)^{k}\varphi ^{i_{1}...i_{d-1}j_{k}}(\xi )\varphi ^{j_{1},...%
\hat{\jmath}_{k}...j_{d+1}}(\xi ),  \tag{83}
\end{equation}%
for $k=1,...,d+1$ there exist $r_{1},...,r_{d+1}\in R^{+}$ such that

\begin{equation}
\tsum \limits_{k=1}^{d+1}r_{k}\omega _{k}(\xi )=0.  \tag{84}
\end{equation}

In the case of a realizable local phirotope we have

\begin{equation}
\Omega ^{i_{1}...i_{d}}(\xi )=\omega (\det (\mathbf{u}^{i_{1}}(\xi ),...,%
\mathbf{u}^{i_{d}}(\xi )),  \tag{85}
\end{equation}%
where $\xi =(\xi ^{1},...,\xi ^{d})$ are local coordinates of some $d$%
-dimensional manifold $B$. The vectors $\mathbf{v}^{i_{1}}(\xi ),...,\mathbf{%
v}^{i_{d}}(\xi )$ can be thought as vectors in the tangent space $T_{\xi
}(B) $ at $\xi $. One can assume that the possibility of considering the
expressions (77)-(79) in a local context may be justified in principle by
the so-called matroid bundle notion (see Refs. [24]-[28]). Let us recall
that the projective variety of decomposable forms is isomorphic to the
Grassmann variety of $d$-dimensional linear subspaces in $R^{n}$. In turn,
the Grassmann variety is the classifying space for vector bundle structures.
Taking these ideas as a motivation, MacPherson developed the combinatorial
differential manifold concept. The matroid bundle notion arises as a
generalization of the MacPherson proposal. Roughly speaking, a matroid
bundle is a structure in which at each point of the differentiable manifold
an oriented matroid is attached as a fiber (see [24]-[28] for details).

Now, let us consider a supermanifold $\mathcal{B}$ parametrized by the local
coordinates $(\xi ,\theta )$ where $\theta $ are elements of the odd
Grassmann algebra (anticommuting variables). We shall now consider the
supersymmetric prescription

\begin{equation}
\mathbf{v}^{i}\rightarrow \mathbf{\pi }^{i}=\mathbf{v}^{i_{1}}-i\bar{\theta}%
\gamma ^{i}\mathbf{\partial }\theta .  \tag{86}
\end{equation}%
Here, $\gamma ^{i}$ are elements of a Clifford algebra. Using (86) one can
generalize (85) in the form

\begin{equation}
\Psi ^{i_{1}...i_{d}}(\xi ,\theta )=\omega (\det (\mathbf{\pi }^{i_{1}}(\xi
,\theta ),...,\mathbf{\pi }^{i_{d}}(\xi ,\theta )).  \tag{87}
\end{equation}%
The symbol $\det $ means the superdeterminant. One should expect that (87)
satisfies a kind of supersymmetric Grassmann-Pl\"{u}cker relations. It is
not difficult to see that up to total derivative (87) is invariant under the
global supersymmetric transformations

\begin{equation}
\delta \theta =\epsilon  \tag{88}
\end{equation}%
and

\begin{equation}
\delta \mathbf{v}^{i_{1}}=i\bar{\epsilon}\gamma ^{i}\partial \theta , 
\tag{89}
\end{equation}%
where $\epsilon $ is a constant complex spinor parameter.

Similarly, one can generalize the superphirotope to the non-representable
case by assuming that if

\begin{equation}
\omega _{k}=(-1)^{k}\varphi ^{i_{1}...i_{d-1}j_{k}}(\xi ,\theta )\varphi
^{j_{1}...\hat{\jmath}_{k}....j_{d+1}}(\xi ,\theta ),  \tag{90}
\end{equation}%
for $k=1,...,d+1$ there exist $r_{1},...,r_{d+1}\in R^{+}$ such that

\begin{equation}
\tsum \limits_{k=1}^{d+1}r_{k}\omega _{k}(\xi ,\theta )=0.  \tag{91}
\end{equation}%
Of course, in the case that the complex structure is projected to the real
structure one should expect that the superphirotope is reduced to the
superchirotope.

With the superphirotope $\Psi ^{i_{1}...i_{d}}(x,\theta )$ at hand one may
consider a possible partition function%
\begin{equation}
Z=\int D\Psi \exp (iS),  \tag{92}
\end{equation}%
where

\begin{equation}
S=\frac{1}{2}\int d^{d}\xi d\theta (\lambda ^{-1}\Psi ^{i_{1}...i_{d}}(\xi
,\theta )\Psi _{^{i_{1}...i_{d}}}(\xi ,\theta )-\lambda T_{d}^{2})  \tag{93}
\end{equation}%
is a Schild type action for a superphirotope. Here, $\lambda $ is a Lagrange
multiplier and $T_{d}$ is the $(d-1)$-phirotope tension. Moreover, in a more
general context the action may have the form

\begin{equation}
S=\frac{1}{2}\int d^{d}\xi d\theta (\lambda ^{-1}\varphi
^{_{^{i_{1}....i_{d}}}}(\xi ,\theta )\varphi _{^{i_{1}...i_{d}}}(\xi ,\theta
)-\lambda T_{d}^{2}).  \tag{94}
\end{equation}%
The advantage of the actions (93) and (94) is that duality is automatically
assured. In fact, in the oriented matroid theory duality is a main subject
in the sense that any chirotope has an associated dual chirotope. This means
that a theory described in the context of an oriented matroid automatically
contains a duality symmetry. Therefore, with our prescription one is
assuring not only the supersymmetry for the action (93) or (94) but also the
duality symmetry.

The action (93) can be related to an ordinary super $p$-brane by assuming
that $\Psi ^{i_{1}...i_{d}}(\xi ,\theta )$ is a closed $d$-form because in
that case we can write

\begin{equation}
\pi _{a}^{i}=\partial _{a}x^{i}-i\bar{\theta}\gamma ^{i}\partial _{a}\theta .
\tag{95}
\end{equation}%
The coordinates $x^{i}$ are the $p$-brane bosonic coordinates. It is worth
mentioning that the bosonic sector of $\Psi ^{i_{1}...i_{d}}(\xi ,\theta )$
is a constraint of the Nambu-Poisson geometry which has been related to
oriented matroid theory (see Ref. [29] for details).

It may be interesting for further research to consider the action (93) from
the point of view of a superfield formalism instead of using the
prescription (95). In this case one may consider a supersymmetrization in
the form $\pi _{a}^{i}(\xi ,\theta )=\partial _{a}X^{i}$, with $X^{i}$ as a
scalar superfield admitting a finite expansion in terms of $\theta .$ For
instance, in four dimensions one may have

\begin{equation}
X^{i}(\xi ,\theta )=x^{i}(\xi )+i\theta \psi ^{i}(\xi )+\frac{i}{2}\bar{%
\theta}\theta B^{i}(\xi ).  \tag{96}
\end{equation}%
The state $\psi ^{i}$ denotes a Majorana spinor field, while $B^{i}$ refers
to an auxiliary field. By substituting (96) into (93) one should expect a
splitting of (93) in several terms containing the variables $x^{i}(\xi
),\psi ^{i}(\xi )$ and $B^{i}(\xi )$. The important thing is that using the
prescription (96) supersymmetry becomes evident in the sense that the
algebra of supersymmetry transformations is closed off the mass-shell.

Although in section $3$ we focused on $11$-dimensional supergravity similar
arguments can be applied to the case of $10$-dimensional supergravity.
Specifically, as we already mentioning by studying maximal supersymmetry in
IIB supergravity Figueroa-O'Farril and Papadopoulos [14]-[15] used the
vanishing of the curvature of the supercovariant derivative to derive the
analogue Grassmann-Pl\"{u}cker formula

\begin{equation}
F_{LP_{1}P_{2}P_{3}[P_{4}}F_{Q_{1}Q_{2}Q_{3}Q_{4}]}^{L}=0,  \tag{97}
\end{equation}%
for the five-form $F_{LP_{1}P_{2}P_{3}P_{4}}$. Moreover, in Refs. [14]-[15]
is proved that (97) implies that

\begin{equation}
F=G+^{\ast }G,  \tag{98}
\end{equation}%
where $G$ is a decomposable $5$-form and $^{\ast }G$ denotes the $10$%
-dimensional dual of $G.$ This means that $G$ and $^{\ast }G$ satisfy the
Grassmann-Pl\"{u}cker relations and therefore can be identified with a $5$%
-rank chirotope.

\bigskip \ 

\noindent \textbf{9. Connection with qubit theory}

\smallskip \ 

A connection between $4$-rebits (real qubits) and the Nambu-Goto action with
target `spacetime' of four time and four space dimensions ($(4+4)$%
-dimensions)) was proposed in Ref. [5]. The motivation for this proposal
came three observations. The first one is that a $4$-rebit contains exactly
the same number of degree of freedom as a complex $3$-qubit and therefore $4$%
-rebits are special in the sense of division algebras. Secondly, the $(4+4)$%
-dimensions can be splitted as $(4+4)=(3+1)+(1+3)$ and therefore they are
connected with an ordinary $(1+3)$-spacetime and with changed signature
associated with $(3+1)$-spacetime [30]. Moreover it was shown how geometric
aspects of $4$-rebits can be related to the chirotope concept of oriented
matroid theory (see Ref. [4]).

It is worth mentioning that the discovery of new hidden discrete symmetries
of the Nambu-Goto action (through the identification of the coordinates $%
x^{\mu }$ of a bosonic string, in target space of $(2+2)$-signature, with a $%
2\times 2$ matrix $x^{ab})$ [31] leads to increase the interest in qubit
theory. It turns out that the key mathematical tool in this development is
the Cayley hyperdeterminant $Det(b)$\ [32] of the hypermatrix $%
b_{a}^{~~bc}=\partial _{a}x^{bc}$. A striking result is that $Det(b)$ can
also be associated with the four electric charges and four magnetic charges
of a STU black hole in four dimensional string theory [33] (see also Ref.
[34]). Even more surprising is the fact that $Det(b)$ makes also its
appearance in quantum information theory by identifying $b_{a}^{~~bc}$ with
a complex $3$-qubit system $a_{a}^{~~bc}$ [35]. These coincidences, among
others, have increased the interest on the qubit/black hole correspondence
[36]-[37].

Additional motivation concerning a connection between the $(4+4)$-signature
and qubit theory may arise from the following observation that $(4+4)$%
-dimensions can also be understood as $(4+4)=((2+2)+(2+2))$. The importance
of the signature $(2+2)$ appears in different physical scenarios, including $%
N=2$ strings (see Ref. [38] and references therein).

It turns out that in information theory $4$-qubit is just subclass of $N$%
-qubit entanglement. In fact, the Hilbert space can be broken into the form $%
C^{2^{N}}=C^{L}\otimes C^{l}$, with $L=2^{N-1}$ and $l=2$. Such a partition
it allows a geometric interpretation in terms of the complex Grassmannian
variety $Gr(L,l)$ of $2$-planes in $C^{L}$ via the Pl\"{u}cker embedding. In
this case, the Pl\"{u}cker coordinates of Grassmannians $Gr(L,l)$ are
natural invariants of the theory (see Refs [19] and [39] for details).

However, in this context, it has been mentioned in Ref. [40], and proved in
Refs. [41] and [42], that for normalized qubits the complex $1$-qubit, $2$%
-qubit and $3$-qubit are deeply related to division algebras via the Hopf
maps, $S^{3}\overset{S^{1}}{\longrightarrow }S^{2}$, $S^{7}\overset{S^{3}}{%
\longrightarrow }S^{4}$ and $S^{15}\overset{S^{7}}{\longrightarrow }S^{8}$,
respectively.

Consider the general complex state $\mid \psi >\in C^{2^{N}},$

\begin{equation}
\mid \psi >=\dsum
\limits_{a_{1},a_{2},...,a_{N}=0}^{1}a_{a_{1}a_{2}...a_{N}}\mid
a_{1}a_{2}...a_{N}>,  \tag{99}
\end{equation}%
where the states $\mid a_{1}a_{2}...a_{N}>=\mid a_{1}>\otimes \mid
a_{2}>...\otimes \mid a_{N}>$ correspond to a standard basis of the $N$%
-qubit. For a $3$-qubit (99) becomes

\begin{equation}
\mid \psi >=\dsum \limits_{a_{1},a_{2},a_{3}=0}^{1}a_{a_{1}a_{2}a_{3}}\mid
a_{1}a_{2}a_{3}>,  \tag{100}
\end{equation}%
while for $4$-qubit one has

\begin{equation}
\mid \psi >=\dsum
\limits_{a_{1},a_{2},a_{3},a_{4}=0}^{1}a_{a_{1}a_{2}a_{3}a_{4}}\mid
a_{1}a_{2}a_{3}a_{4}>.  \tag{101}
\end{equation}%
It is interesting to make the following observations. First, one notes that $%
a_{a_{1}a_{2}a_{3}}$ has $8$ complex degrees of freedom, that is $16$ real
degrees of freedom, while $a_{a_{1}a_{2}a_{3}a_{4}}$ contains $16$ complex
degrees of freedom, that is $32$ real degrees of freedom. Let us denote $N$%
-rebit system (real $N$-qubit ) by $b_{a_{1}a_{2}...a_{N}}$. So we shall
denote the corresponding $3$-rebit, $4$-rebit by $b_{a_{1}a_{2}a_{3}}$ and $%
b_{a_{1}a_{2}a_{3}a_{4}}$, respectively. One observes that $%
b_{a_{1}a_{2}a_{3}}$ has $8$ real degrees of freedom, while $%
b_{a_{1}a_{2}a_{3}a_{4}}$ has $16$ real degrees of freedom. Thus, by this
simple (degree of freedom) counting one note that it seems more natural to
associate the $4$-rebit $b_{a_{1}a_{2}a_{3}a_{4}}$ with the complex $3$%
-qubit, $a_{a_{1}a_{2}a_{3}}$, than with the complex $4$-qubit, $%
a_{a_{1}a_{2}a_{3}a_{4}}$. Of course, by imposing some constraints one can
always reduce the $32$ real degrees of freedom of $a_{a_{1}a_{2}a_{3}a_{4}}$
to $16$, and this is the kind of embedding discussed in Ref. [19]. The main
idea in reference [5] was to make sense out of a $4$-rebit in the Nambu-Goto
context without loosing the important connection with a division algebra via
the Hopf map $S^{15}\overset{S^{7}}{\longrightarrow }S^{8}$.

Let us first show the formalism concerning the Nambu-Goto action/qubits
correspondence in a spacetime of $(2+2)$-signature. In the $(2+2)$%
-dimensions one may introduce the matrix%
\begin{equation}
x^{ab}=\left( 
\begin{array}{cc}
x^{1}+x^{3} & x^{2}+x^{4} \\ 
x^{2}-x^{4} & -x^{1}+x^{3}%
\end{array}%
\right) .  \tag{102}
\end{equation}%
Using (102) the line element%
\begin{equation}
ds^{2}=dx^{\mu }dx^{\nu }\eta _{\mu \nu },  \tag{103}
\end{equation}%
can also be written as

\begin{equation}
ds^{2}=\frac{1}{2}dx^{ab}dx^{cd}\varepsilon _{ac}\varepsilon _{bd}, 
\tag{104}
\end{equation}%
where%
\begin{equation}
\eta _{\mu \nu }=diag(-1,-1,1,1),  \tag{105}
\end{equation}%
is a flat metric corresponding to $(2+2)$-signature and $\varepsilon _{ab}$
is the completely antisymmetric symbol ($\varepsilon $-symbol) with $%
\varepsilon _{12}=1$.

On the other hand in a target space of $(4+4)$-signature one may introduce
the matrices

\begin{equation}
x^{ab1}=\left( 
\begin{array}{cc}
x^{1}+x^{5} & x^{2}+x^{6} \\ 
x^{2}-x^{6} & -x^{1}+x^{5}%
\end{array}%
\right) ,  \tag{106}
\end{equation}%
and%
\begin{equation}
x^{ab2}=\left( 
\begin{array}{cc}
x^{3}+x^{7} & x^{4}+x^{8} \\ 
x^{4}-x^{8} & -x^{3}+x^{7}%
\end{array}%
\right) .  \tag{107}
\end{equation}

At first sight one may consider the line element

\begin{equation}
ds^{2}=\frac{1}{2}dx^{abc}dx^{def}\varepsilon _{ad}\varepsilon
_{be}\varepsilon _{cf}  \tag{108}
\end{equation}%
as the analogue of (104). But this vanishes identically because 
\begin{equation}
s^{cf}\equiv dx^{abc}dx^{def}\varepsilon _{ad}\varepsilon _{be}  \tag{109}
\end{equation}%
is a symmetric quantity, while $\varepsilon _{cf}$ is antisymmetric. In
fact, the correct line element in $(4+4)$-dimensions turns out to be

\begin{equation}
ds^{2}=\frac{1}{2}dx^{abc}dx^{def}\varepsilon _{ad}\varepsilon _{be}\eta
_{cf}.  \tag{110}
\end{equation}%
Notice that we have changed the last $\varepsilon $-symbol in (110) for the $%
\eta $-symbol. Here, $\eta _{cf}=diag(-1,1)$. Moreover, one can prove that
(103), with 
\begin{equation}
\eta _{\mu \nu }=(-1,-1,-1,-1,+1,+1,+1,+1),  \tag{111}
\end{equation}%
follows from (110).

Similarly, the world sheet metric in $(2+2)$-dimensions

\begin{equation}
\gamma _{ab}=\partial _{a}x^{\mu }\partial _{b}x^{\nu }\eta _{\mu \nu
}=\gamma _{ba},  \tag{112}
\end{equation}%
can be written as%
\begin{equation}
\gamma _{ab}=\frac{1}{2}\partial _{a}x^{cd}\partial _{b}x^{ef}\varepsilon
_{ce}\varepsilon _{df}.  \tag{113}
\end{equation}%
While in $(4+4)$-dimensions, one has%
\begin{equation}
\gamma _{ab}=\frac{1}{2}b_{a}^{~~cdg}b_{b}^{~~fhl}\varepsilon
_{cf}\varepsilon _{dh}\eta _{gl},  \tag{114}
\end{equation}%
with

\begin{equation}
b_{a}^{~~cdg}\equiv \partial _{a}x^{cdg}.  \tag{115}
\end{equation}

In $(2+2)$-dimensions one can write the determinant of $\gamma _{ab}$,

\begin{equation}
\det \gamma =\frac{1}{2}\varepsilon ^{ab}\varepsilon ^{cd}\gamma _{ac}\gamma
_{bd},  \tag{116}
\end{equation}%
in the hyperdeterminant form

\begin{equation}
\det \gamma =\frac{1}{2}\varepsilon ^{ab}\varepsilon ^{cd}\varepsilon
_{eg}\varepsilon _{fh}\varepsilon _{ru}\varepsilon
_{sv}b_{a}^{~~ef}b_{c}^{~~gh}b_{b}^{~~rs}b_{d}^{~~uv}=Det(b),  \tag{117}
\end{equation}%
with

\begin{equation}
b_{a}^{~~cd}\equiv \partial _{a}x^{cd}.  \tag{118}
\end{equation}%
Thus, this proves that the Nambu-Goto action

\begin{equation}
S=\frac{1}{2}\int d^{2}\xi \sqrt{\det \gamma },  \tag{119}
\end{equation}%
for a flat target \textquotedblleft spacetime\textquotedblright \ with $%
(2+2) $-signature can also be written as [31]

\begin{equation}
S=\frac{1}{2}\int d^{2}\xi \sqrt{Det(b)}.  \tag{120}
\end{equation}

While in $(4+4)$-dimensions the determinant

\begin{equation}
\det \gamma =\frac{1}{2}\varepsilon ^{ab}\varepsilon ^{cd}\gamma _{ac}\gamma
_{bd},  \tag{121}
\end{equation}%
can be written as

\begin{equation}
\det \gamma =\frac{1}{2}c^{efrs}c^{ghuv}\varepsilon _{eg}\varepsilon
_{fh}\varepsilon _{ru}\varepsilon _{sv},  \tag{122}
\end{equation}%
where

\begin{equation}
c^{efrs}\equiv (-\varepsilon ^{ab}b_{a}^{~~ef1}b_{b}^{~~rs1}+\varepsilon
^{ab}b_{a}^{~~ef2}b_{b}^{~~rs2}),  \tag{123}
\end{equation}%
One recognizes in (121) the hyperdeterminant of the hypermatrix $c^{efrs}$.
So, we can write

\begin{equation}
\det \gamma =Det(c).  \tag{124}
\end{equation}%
This proves that the Nambu-Goto action in $(4+4)$-dimensions%
\begin{equation}
S=\frac{1}{2}\int d^{2}\xi \sqrt{\det \gamma },  \tag{125}
\end{equation}%
can also be written as [5]

\begin{equation}
S=\frac{1}{2}\int d^{2}\xi \sqrt{Det(c)}.  \tag{126}
\end{equation}

Moreover, one may connect qubits with the chirotope concept in oriented
matroid theory. In space of $(2+2)$-signature one writes

\begin{equation}
\det \gamma =\frac{1}{2}\sigma ^{\mu \nu }\sigma ^{\alpha \beta }\eta _{\mu
\alpha }\eta _{\nu \beta },  \tag{127}
\end{equation}%
where

\begin{equation}
\sigma ^{\mu \nu }=\varepsilon ^{ab}b_{a}^{\mu }b_{b}^{\nu }.  \tag{128}
\end{equation}%
Here, we have used the definition

\begin{equation}
b_{a}^{\mu }\equiv \partial _{a}x^{\mu }.  \tag{129}
\end{equation}%
Since $\sigma ^{\mu \nu }$ satisfies the identity $\sigma ^{\mu \lbrack \nu
}\sigma ^{\alpha \beta ]}\equiv 0$, one can verify that $\chi ^{\mu \nu
}=sign\sigma ^{\mu \nu }$ satisfies the Grassmann-Pl\"{u}cker relation

\begin{equation}
\chi ^{\mu \lbrack \nu }\chi ^{\alpha \beta ]}=0,  \tag{130}
\end{equation}%
and therefore $\chi ^{\mu \nu }$ is a realizable chirotope (see Refs.
[4]-[5] and references therein).

The Grassmann-Pl\"{u}cker relation (130) implies that the ground set is 
\begin{equation}
E=\{ \mathbf{1,2,3,4}\}  \tag{131}
\end{equation}%
and the alternating map%
\begin{equation}
\chi ^{\mu \nu }\rightarrow \{-1,0,1\},  \tag{132}
\end{equation}%
determine a $2$-rank realizable oriented matroid $M=(E,\chi ^{\mu \nu })$.
The collection of bases for this oriented matroid is

\begin{equation}
\mathcal{B}=\{ \mathbf{\{1,2\},\{1,3\},\{1,4\},\{2,3\},\{2,4\},\{3,4\}}\}, 
\tag{133}
\end{equation}%
which can be obtained by just given values to the indices $\mu $ and $\nu $
in $\chi ^{\mu \nu }$. Indeed, the pair $(E,\mathcal{B})$ determines a $2$%
-rank uniform non-oriented ordinary matroid.

In the case of qubits, one may introduce the underlying ground bitset (from
bit and set)%
\begin{equation}
\mathcal{E}=\{1,2\}  \tag{134}
\end{equation}%
and the pre-ground set

\begin{equation}
E_{0}=\{(1,1),(1,2),(2,1),(2,2)\}.  \tag{135}
\end{equation}%
It turns out that $E_{0}$ and $E$ can be related by establishing the
identification

\begin{equation}
\begin{array}{cc}
(1,1)\leftrightarrow \mathbf{1}, & (1,2)\leftrightarrow \mathbf{2}, \\ 
&  \\ 
(2,1)\leftrightarrow \mathbf{3}, & (2,2)\leftrightarrow \mathbf{4}.%
\end{array}
\tag{136}
\end{equation}%
Observe that (136) is equivalent of making the identification of indices $%
\{a,b\} \leftrightarrow \mu $,..,etc. In fact, considering these
identifications the family of bases (133) becomes

\begin{equation}
\begin{array}{c}
\mathcal{B}_{0}=\{ \{(1,1),(1,2)\},\{(1,1),(2,1)\},\{(1,1),(2,2)\}, \\ 
\\ 
\{(1,2),(2,1)\},\{(1,2),(2,2)\},\{(2,1),(2,2)\} \}.%
\end{array}
\tag{137}
\end{equation}

Using the definition

\begin{equation}
\sigma ^{efrs}\equiv \varepsilon ^{ab}b_{a}^{~~ef}b_{b}^{~~rs},  \tag{138}
\end{equation}%
one can show that

\begin{equation}
\det \gamma =\frac{1}{2}\sigma ^{efrs}\sigma ^{ghuv}\varepsilon
_{eg}\varepsilon _{fh}\varepsilon _{ru}\varepsilon _{sv}=Det(b).  \tag{139}
\end{equation}%
This establishes a link between the hyperdeterminant and \textquotedblleft
chirotope\textquotedblright \ structure.

In $(4+4)$-dimensions one may introduce the quantity

\begin{equation}
c^{\mu \nu }=(-\varepsilon ^{ab}b_{a}^{\mu 1}b_{b}^{\nu 1}+\varepsilon
^{ab}b_{a}^{\mu 2}b_{b}^{\nu 2}).  \tag{140}
\end{equation}%
Here, one has considered the definitions%
\begin{equation}
b_{a}^{\mu 1}=\partial _{a}x^{\mu }  \tag{141}
\end{equation}%
and

\begin{equation}
b_{a}^{\mu 2}=\partial _{a}y^{\mu }.  \tag{142}
\end{equation}%
In turn this means that we can write

\begin{equation}
c^{\mu \nu }=(-\varepsilon ^{ab}\partial _{a}x^{\mu }\partial _{b}x^{\nu
}+\varepsilon ^{ab}\partial _{a}y^{\mu }\partial _{b}y^{\nu }).  \tag{143}
\end{equation}%
One recognizes in this expression the Pl\"{u}cker coordinates for both cases 
$u_{a}^{\mu }=\partial _{a}x^{\mu }$ and $v_{a}^{\mu }=\partial _{a}y^{\mu }$%
.

This proves that both quantities $\sigma ^{efrs}$ and $c^{efrs}$
(qubitopes), belongs to an underlying structure $Q=(\mathcal{E},E_{0},B_{0})$
called qubitoid. The word \textquotedblleft qubitoid\textquotedblright \ is
a short word for qubit-matroid.

One may now be interested to see how the $4$-rebit $b_{a_{1}a_{2}a_{3}a_{4}}$
is connected with $a_{a_{1}a_{2}a_{3}}$. The simplest (but no the most
general) possibility seems to be

\begin{equation}
a_{a_{1}a_{2}a_{3}}=b_{a_{1}a_{2}a_{3}1}+ib_{a_{1}a_{2}a_{3}2}.  \tag{144}
\end{equation}%
In turn this implies

\begin{equation}
a_{a_{1}}^{~~a_{2}a_{3}}=\partial _{a_{1}}x^{a_{2}a_{3}1}+i\partial
_{a_{1}}x^{a_{2}a_{3}2}=\partial _{a_{1}}(x^{a_{2}a_{3}1}+ix^{a_{2}a_{3}2}).
\tag{145}
\end{equation}%
Hence the 3-qubit $a_{a_{1}}^{~~a_{2}a_{3}}$ is related to the two $2$%
-rebits states $x^{a_{2}a_{3}1}$ and $x^{a_{2}a_{3}2}$. This observation may
help eventually to relate $4$-rebit $b_{a_{1}a_{2}a_{3}a_{4}}$ with the Hopf
fibration $S^{15}\overset{S^{7}}{\longrightarrow }S^{8}$. In fact, a
normalization of the complex states $a_{a_{1}a_{2}a_{3}}$ leads to the $15$%
-dimensional sphere $S^{15}$ which under the Hopf map, admit parametrization
of the parallelizable sphere $S^{7}$ fibration over $S^{8}$.

It turns out, that just as the norm group of quaternions is $%
SO(4)=S^{3}\times S^{3}$ , the norm group of octonions is $SO(8)=S^{7}\times
S^{7}\times G_{2}$ (see Ref. [22]). This is due to the fact that considering
the 28 generators $J_{\mu \nu }$ of $SO(8)$ and the octonionic $o_{i}$
structure constants $\psi _{ijk}$ ($o_{i}o_{j}=(\psi _{i})_{j}^{k}o_{k}$)
one can choose a basis $M_{i}=J_{0i}$, $K_{i}=\frac{1}{2}\psi _{ijk}J^{jk}$
and $\Gamma _{ij}=2J_{ij}-\frac{1}{3!2}\varepsilon _{ijklmns}\psi
^{mns}J^{kl}$ for $SO(8)$ satisfying the algebra,

\begin{equation}
\begin{array}{c}
\lbrack M_{i},M_{j}]=\frac{1}{3}(\psi _{ijk}K^{k}+\Gamma _{ij}), \\ 
\\ 
\lbrack K_{i},K_{j}]=-\psi _{ijk}K^{k}+\Gamma _{ij}, \\ 
\\ 
\lbrack K_{i},M_{j}]=\psi _{ijk}M^{k}, \\ 
\\ 
\lbrack \Gamma _{ij},\Gamma _{ij}]=C_{ijklms}\Gamma ^{ms}.%
\end{array}
\tag{146}
\end{equation}%
Here,

\begin{equation}
C_{ijklms}=\mathcal{A}\text{(}\frac{3}{2}\delta _{il}\delta _{jm}\delta
_{ls}-\frac{1}{8}\psi _{ijm}\psi _{kls}).  \tag{147}
\end{equation}%
The $\mathcal{A}$ in (147) stands for antisymmetrization of $i$ and $j,$ $k$
and $l$, and $m$ and $s$. The relevant aspect is that from the algebra (146)
one discovers that the operators $M_{i}+K_{i}$ and $M_{i}-K_{i}$ commute and
therefore they corresponds to independent $7$-spheres $S_{R}^{7}$ and $%
S_{L}^{7}$. In this way the decomposition $M_{i}+K_{i}$, $M_{i}-K_{i}$ and $%
L_{ij}$ of the generators of the group $SO(8)$ leads to the decomposition $%
SO(8)=S_{R}^{7}\times S_{L}^{7}\times G_{2}$. Roughly speaking one can say
that the coset $SO(8)/SO(7)$ is associated with $M_{i}$, the coset $%
SO(7)/G_{2}$ with $K_{i}$ and $C_{ijklms}$ determines the structure
constants of $G_{2}$ (see Ref. [22] for details).

Since in the $4+4$-signature the relevant group is $SO(4,4)$. One find that $%
8$-dimensional spinor representation associated with $spin(8)$ can be
written as

\begin{equation}
\left( 
\begin{array}{cc}
0 & (\psi _{i})_{j}^{k} \\ 
-(\psi _{i})_{j}^{k} & 0%
\end{array}%
\right) .  \tag{148}
\end{equation}%
This means that when $SO(8)$ decomposed under the subgroup $SO(4)\times
SO(4) $ one gets irreducible representation

\begin{equation}
8\longrightarrow (4,1)+(1,4).  \tag{149}
\end{equation}%
Thus, in the case of $SO(4,4)$ one may consider decomposition under the
subgroup $SO(2,2)\times SO(2,2)$ obtaining,

\begin{equation}
(4+4)\longrightarrow ((2+2),1)+(1,(2+2)).  \tag{150}
\end{equation}%
It turns out that these two direct summands correspond to the variables $%
x^{ab1}$ and $x^{ab2}$. This explains why $dx^{abc}$, is contracted with $%
\eta _{ab}$, and no with $\varepsilon _{ab}$.

The above scenario can be generalized for class of $N$-qubits, with the
Hilbert space in the form $C^{2^{N}}=C^{L}\otimes C^{l}$, with $L=2^{N-n}$
and $l=2^{n}$. Such a partition allows a geometric interpretation in terms
of the complex Grassmannian variety $Gr(L,l)$ of $l$-planes in $C^{L}$ via
the Pl\"{u}cker embedding [19]. In the case of $N$-rebits one can set a $%
L\times l$ matrix variable $b_{a}^{\mu }$, $\mu =1,2,...,L$, $a=1,2...,l$,
of $2^{N}=L\times l$ associated with the variable $b_{a_{1}a_{2}...a_{N}}$,
with $a_{1},a_{2,...}etc$ taking values in the set $\{1,2\}$. In fact, one
can take the first $N-n$ terms in $b_{a_{1}a_{2}...a_{N}}$ are represented
by the index $\mu $ in $b_{a}^{\mu }$, while the remaining $n$ terms are
considered by the index $a$ in $b_{a}^{\mu }$. One of the advantage of this
construction is that the Pl\"{u}cker coordinates associated with the real
Grassmannians $b_{a}^{\mu }$ are natural invariants of the theory. Since
oriented matroid theory leads to the chirotope concept which is also defined
in terms Pl\"{u}cker coordinates these developments establishes a possible
link between chirotopes, qubitoids and $p$-branes.

Moreover, since it has been shown [43] that the Duff's discovery of manifest 
$SL(2,R)\times SL(2,R)\times SL(2,R)$ symmetry of the Nambu-Goto action can
be extended to the Green-Schwarz $N=2$ string action it seems interesting to
see whether the developments presented in this section may provide a useful
mathematical tools in this context. The central observation in this case is
that the Cayley's hyperdeterminant in the supersymmetric system is also
related to ordinary determinant in the form%
\begin{equation}
\mathcal{D}et(\Pi _{i\alpha \dot{\beta}})=\epsilon ^{ij}\epsilon
^{kl}\epsilon ^{\alpha \beta }\epsilon ^{\gamma \delta }\epsilon ^{\dot{%
\alpha}\dot{\beta}}\epsilon ^{\dot{\gamma}\dot{\delta}}\Pi _{i\alpha \dot{%
\alpha}}\Pi _{i\beta \dot{\beta}}\Pi _{i\gamma \dot{\gamma}}\Pi _{i\delta 
\dot{\delta}}=\det (\Sigma _{ij}),  \tag{151}
\end{equation}%
with%
\begin{equation}
\Sigma _{ij}=\eta _{\bar{a}\bar{b}}\Pi _{i}^{\bar{a}}\Pi _{j}^{\bar{b}}. 
\tag{152}
\end{equation}%
Here, $\eta _{\bar{a}\bar{b}}=diag(-,-,+,+)$ is the$(2+2)$-dimensional flat
space-time metric and $\Pi _{i}^{\bar{a}}=(\partial _{i}Z^{M})E_{M}^{\bar{a}%
} $, with the target superspace coordinates $Z^{M}$ (see Ref. [43] for
details). This means that it must be possible to make the expression (117),
and therefore the action (119), supersymmetric. In turn this may motivate to
consider supersymmetric aspects in the context of qubit theory for a
space-time in $(4+4)$-dimensions.

\bigskip \ 

\noindent \textbf{10. Final remarks}

\smallskip \ 

It is evident from the discussion of the previous sections that the
Grassmann-Pl\"{u}cker relations play a central role on a number of links
between different physical and mathematical scenarios including Grassmannian
varieties, $11$-dimensional supergravity, qubit theory, $p$-branes and
oriented matroid theory. If a $p$-form satisfies the Grassmann-Pl\"{u}cker
relations then such a $p$-form is decomposable. An application of this
result to maximally supersymmetric solutions of $11$-dimensional
supergravity opens the possibility for writing the $4$-form field strength $%
F_{\hat{\mu}_{1}\hat{\mu}_{2}\hat{\mu}_{3}\hat{\mu}_{4}}$ in terms of some
kind of gauge field $F_{\hat{\mu}}^{a}$. Thus, following this kind of though
one is lead to look for the analogue $F_{\hat{\mu}}^{a}$ for the Bianchi
identities $dF=0$ associated with $F_{\hat{\mu}_{1}\hat{\mu}_{2}\hat{\mu}_{3}%
\hat{\mu}_{4}}$. In this case, it is found that $F_{\hat{\mu}_{1}\hat{\mu}%
_{2}\hat{\mu}_{3}\hat{\mu}_{4}}$ can be written in terms of a gauge field $%
A_{\hat{\mu}_{1}\hat{\mu}_{2}\hat{\mu}_{3}}$ in the form $F_{\hat{\mu}_{1}%
\hat{\mu}_{2}\hat{\mu}_{3}\hat{\mu}_{4}}=\partial _{\lbrack \hat{\mu}_{1}}A_{%
\hat{\mu}_{2}\hat{\mu}_{3}\hat{\mu}_{4}]}$. So, the question arises whether
the Bianchi identities and the Grassmann-Pl\"{u}cker relations associated
with a general $p$-form field strength $F_{\mu _{1}...\mu _{p}}$ are
connected. In other word, the challenge is to know what is the relation
between $F_{\hat{\mu}_{1}}^{a_{1}}$ and $A_{\mu _{1}...\mu _{p-1}}$.

Let us assume that a local $p$-form $F_{\mu _{1}...\mu _{p}}$ satisfies the
Grassmann-Plucker relations

\begin{equation}
F_{\mu _{1}...[\mu _{p}}F_{\nu _{1}...\nu _{p}]}=0.  \tag{153}
\end{equation}%
According to our previous discussion one knows that (153) implies that $%
F_{\mu _{1}...\mu _{p}}$ is decomposable. This means that $F_{\mu _{1}...\mu
_{p}}$ can be written as

\begin{equation}
F_{\mu _{1}...\mu _{p}}=\varepsilon _{a_{1}...a_{p}}F_{\mu
_{1}}^{a_{1}}...F_{\mu _{p}}^{a_{p}}.  \tag{154}
\end{equation}

Now, let us assume that $dF=0$. In tensorial notation this means

\begin{equation}
\partial _{\mu _{p+1}}F_{\mu _{1}...\mu _{p}}dx^{\mu _{1}}\wedge ...\wedge
dx^{\mu _{p}}\wedge dx^{\mu _{p+1}}=0.  \tag{155}
\end{equation}%
But from (154) one obtains

\begin{equation}
\varepsilon _{a_{1}...a_{p}}\partial _{\mu _{p+1}}(F_{\mu
_{1}}^{a_{1}}...F_{\mu _{p}}^{a_{p}})dx^{\mu _{1}}\wedge ...\wedge dx^{\mu
_{p}}\wedge dx^{\mu _{p+1}}=0,  \tag{156}
\end{equation}%
which implies

\begin{equation}
\varepsilon _{a_{1}...a_{p}}\partial _{\mu _{p+1}}(F_{\mu
_{1}}^{a_{1}})F_{\mu _{2}}^{a_{2}}...F_{\mu _{p}}^{a_{p}}dx^{\mu _{1}}\wedge
...\wedge dx^{\mu _{p}}\wedge dx^{\mu _{p+1}}=0.  \tag{157}
\end{equation}%
So, one discovers that a solution of (157) is given by

\begin{equation}
F_{\mu }^{a}=\frac{\partial \lambda ^{a}}{\partial x^{\mu }}.  \tag{158}
\end{equation}%
Consequently, the expression (154) becomes%
\begin{equation}
F_{\mu _{1}...\mu _{p}}=\varepsilon _{a_{1}...a_{p}}\frac{\partial \lambda
^{a_{1}}}{\partial x^{\mu _{1}}}...\frac{\partial \lambda ^{a_{p}}}{\partial
x^{\mu _{p}}}.  \tag{159}
\end{equation}%
In turn this expression leads to

\begin{equation}
F^{\mu _{1}...\mu _{p}}=\varepsilon ^{a_{1}...a_{p}}\frac{\partial x^{\mu
_{1}}}{\partial \lambda ^{a_{1}}}...\frac{\partial x^{\mu _{p}}}{\partial
\lambda ^{a_{p}}}.  \tag{160}
\end{equation}

On the other hand (155) implies that

\begin{equation}
F_{\mu _{1}...\mu _{p}}=\partial _{\lbrack \mu _{p}}A_{\mu _{1}...\mu
_{p-1}]}.  \tag{161}
\end{equation}%
Hence, using (159) one obtains

\begin{equation}
A_{\mu _{1}...\mu _{p-1}}=\varepsilon _{a_{1}...a_{p}}\lambda ^{a_{p}}\frac{%
\partial \lambda ^{a_{1}}}{\partial x^{\mu _{1}}}...\frac{\partial \lambda
^{a_{p-1}}}{\partial x^{\mu _{p-1}}}.  \tag{162}
\end{equation}%
This establishes a connection between the Bianchi identity $dF=0$ and the
Grassmann-Pl\"{u}cker relations. It is worth mentioning that the expression
(160) is a key tool in the $p$-brane theory, when one writes the Nambu-Goto
action for $p$-branes in Schild type formalism (see Ref. [3] for details).

\bigskip \ 

\textbf{Acknowledgment: }I would like to thank the hospitality of the
Mathematical, Computational and Modeling Sciences Center at the Arizona
State University where part of this work was developed.

\end{document}